\documentclass[11pt]{article}
\usepackage[pdftex]{graphicx,color} 
\usepackage{jheppub}
\usepackage{amsmath}
\usepackage{amssymb}
\usepackage{comment}
\usepackage{multirow}
\usepackage{mathtools}
\usepackage{dsdshorthand}
\usepackage{tikz}
\usepackage{shuffle}
\usetikzlibrary{arrows}
\tikzset{
>=stealth'
}
\newcommand{\bea}{\begin{equation}\begin{aligned}}
\newcommand{\eea}[1]{\label{#1}\end{aligned}\end{equation}}
\newcommand{\beq}{\begin{equation}}
\newcommand{\eeq}{\end{equation}}

\title{Stringy KLT Relations on $AdS$}
\author{Luis F. Alday,}
\author{Maria Nocchi and}
\author{Aurélie Strömholm Sangaré}
\affiliation{Mathematical Institute, University of Oxford, Andrew Wiles Building, Radcliffe Observatory Quarter, Woodstock Road, Oxford, OX2 6GG, U.K.}

\abstract{ 
 We study the building blocks of open and closed string amplitudes on $AdS$. These are given by two infinite towers of world-sheet integrals generalising the Euler and complex beta functions respectively. We show that the open and closed building blocks are related by an $AdS$ version of the KLT relations, whose Kernel can be computed exactly. We furthermore show that the building blocks for open string amplitudes are given by Aomoto-Gelfand hypergeometric functions, and give their closed-form expression up to weight four.}

\emailAdd{\{alday,Maria.Nocchi,Aurelie.Sangare\}@maths.ox.ac.uk}

\begin{document}

\maketitle
\section{Motivation and summary}

The aim of this paper is to study the mathematical structure and properties of the functions that appear in the scattering of open and closed strings on $AdS$ backgrounds. Let us start by reviewing the relevant results in flat space. The Veneziano amplitude for the scattering of four open-string tachyons is given by the Euler beta function
\begin{equation}
\beta(s,t) = \int_0^1 x^{s-1}(1-x)^{t-1} dx = \frac{\Gamma(s)\Gamma(t)}{\Gamma(s+t)} \hspace{2pt}.
\end{equation}
While the integral converges for $\Re (s)>0, \Re (t)>0$, the right-hand side can be continued beyond this region. The closed string amplitude for the scattering of four tachyons in flat space, the Virasoro-Shapiro amplitude, is given by the complex beta function
\begin{equation}
\beta_{\mathbb{C}}(s,t) = \int |z|^{2s-2}|1-z|^{2t-2} d^2 z = \frac{\Gamma(s)\Gamma(t)\Gamma(1-s-t)}{\Gamma(s+t)\Gamma(1-s)\Gamma(1-t)} \hspace{2pt}.
\end{equation}
Again, while the integral converges for $\Re(s)>0,\Re(t)>0,\Re(1-s-t)>0$, the right-hand side can be continued beyond this region. These two functions satisfy various functional equations and relations. For instance, under shifts in the Mandelstam variables
\begin{equation}
\beta(s+1,t) =\frac{s}{s+t} \beta(s,t),~~~\beta(s,t+1) =\frac{t}{s+t} \beta(s,t) \hspace{2pt}.
\end{equation}
Another functional equation is 
\begin{equation}
\label{Poincare}
\beta(s,t)\beta(-s,-t)  =-\frac{\pi  (s+t) (\cot (\pi  s)+\cot (\pi  t))}{s t} \hspace{2pt},
\end{equation}
which can be shown to be related to Poincaré duality \cite{Brown:2019wna}. The Euler and complex beta functions are intimately connected. In a small $s,t$ expansion they are related by a single-valued map
\begin{equation}
\beta_{\mathbb{C}}(s,t) = sv(\beta(s,t))
\end{equation}
where $sv$ acts term by term and is such that
\begin{equation}
sv(\zeta(2n))=\zeta_{sv}(2n)=0,~~~sv(\zeta(2n+1))=\zeta_{sv}(2n+1)=2\zeta(2n+1),~~~n=1,2,\cdots.
\end{equation}
The implications of this relation away from $s=t=0$ are not clear, but it can be shown  \cite{Brown:2018omk,Brown:2019wna} that this is equivalent to
\begin{equation}
\label{KLT}
\beta_{\mathbb{C}}(s,t) =\frac{\sin (\pi  s) \sin (\pi  t) }{\pi \sin (\pi  (s+t))} \beta(s,t)^2,
\end{equation}
which are the KLT relations \cite{Kawai:1985xq}. The relation between open and closed tree-level string amplitudes in flat space goes beyond four points. It is now understood that closed string amplitudes, of any multiplicity, are the single-valued projection of open string amplitudes. This was initially conjectured in  \cite{Stieberger:2013wea} and then shown in \cite{Schlotterer:2018zce,Vanhove:2018elu,Brown:2019wna,Vanhove:2020qtt} using different methods. See also \cite{Stieberger:2014hba,Stieberger:2016xhs,Mafra:2022wml}. 

Developments in flat space are based on explicit results obtained via world-sheet computations. This leads to integral representations for tree-level scattering amplitudes, over the boundary of the disk for open strings, and over the Riemann sphere for closed strings.  For curved spaces in the presence of RR-fluxes, we do not currently have a world-sheet theory capable of computing scattering amplitudes. However, in the presence of $AdS$ factors, on-shell string scattering amplitudes can be defined in terms of correlators of local operators in the CFT at the boundary. CFT tools can then be combined with ideas from number theory to advance the computation of massless tree-level amplitudes for both closed \cite{Alday:2022uxp,Alday:2022xwz,Alday:2023jdk,Alday:2023mvu,Fardelli:2023fyq,Chester:2024wnb,Chester:2024esn} and open \cite{Alday:2024yax,Alday:2024ksp} strings. In a small curvature expansion,\footnote{More precisely, there is a sequence of integral transforms from the position-space four-point correlator to the world-sheet integrals studied here. This notably involves the standard Mellin transform followed by a Borel transform, in the spirit of \cite{Penedones:2010ue}. See, for example, \cite{Alday:2023mvu} for further details.}
the results admit an integral representation, over the boundary of the disk for open strings, and over the Riemann sphere for closed strings, exactly as in flat space, but with the extra insertion of polylogarithms: multiple polylogarithms (MPLs) in the open string case,  single-valued multiple polylogarithms (SVMPLs) in the closed string case. In particular, the low-energy expansion (small $s,t$) then involves only zeta values (open strings) and single-valued zeta values (closed strings), mirroring the expansions in flat space. An analogous structure has also been observed for scattering amplitudes in $AdS_3$ with pure NS–NS fluxes \cite{Alday:2024rjs}. Altogether, these results suggest a certain universality in the structure of string amplitudes on curved backgrounds, which motivates us to consider the natural generalisations of the Euler and complex beta functions
\begin{align}
\begin{split}
J_w(s,t) &= \int_0^1 x^{s-1}(1-x)^{t-1} L_w(x) dx \hspace{2pt}, \\
I_w(s,t) &=  \int |z|^{2s-2}|1-z|^{2t-2} {\cal L}_w(z) d^2 z \hspace{2pt},
\end{split}
\end{align}
where $L_w(x)$ is the MPL labelled by the word $w$ and  ${\cal L}_w(z)$ is the SVMPL labelled by the word $w$. For the empty word $w=e$, we obtain back the Euler and complex beta functions
\begin{equation}
J_e(s,t)= \beta(s,t) \hspace{2pt},~~~I_e(s,t)= \beta_{\mathbb{C}}(s,t).
\end{equation}
We show that these building blocks satisfy several properties and relations. The first key result is that the $J_w(s,t)$  integrals are a particular class of Aomoto-Gelfand hypergeometric functions. Below we give explicit results up to weight four. The second key result is that the integrals $I_w(s,t)$, the building blocks of closed string amplitudes on $AdS$, can be written as bilinears of the integrals $J_w(s,t)$, the building blocks of open string amplitudes on $AdS$:
\begin{equation}
{\cal I}(s,t; e_0,e_1) =  {\cal J}(s,t; e_0,e_1) {\cal K}(s,t;e_0,e_1)  {\cal J}^R(s,t; e_0,e'_1)
\end{equation}
where $e_0,e_1$ are non-commutative variables, $e'_1$ a certain deformation of $e_1$, and $ {\cal I}(s,t; e_0,e_1)$ and ${\cal J}(s,t; e_0,e_1) $ are generating functions for the integrals in question.  This generalises the stringy KLT relations (\ref{KLT}) to scattering on $AdS$. The $AdS$ KLT Kernel takes a remarkably simple form 
\begin{eqnarray}
 {\cal K}(s,t;e_0,e_1)= -\frac{1}{2\pi i} \left(1+\frac{e^{2 i \pi s} M_0}{1-e^{2 i \pi s} M_0}+\frac{e^{2 i \pi t} M_1}{1-e^{2 i \pi t} M_1}\right)^{-1}
\end{eqnarray}
where $M_0,M_1$ are monodromy operators introduced below. The rest of the paper is organised as follows. In section \ref{open}, we focus on the building blocks of open string amplitudes, discuss their properties in general and compute them explicitly up to weight four.  In section \ref{closed}, we focus on the building blocks of closed string amplitudes, and show their precise relation to the open building blocks, providing KLT relations for string scattering on $AdS$. We end with some conclusions. Appendix \ref{conventions} contains a short review of MPLs and SVMPLs. Finally, appendices \ref{poles} and \ref{poincare} discuss additional properties of the $J_w(s,t)$ integrals, including their pole structure and the implications of Poincaré duality. 

\section{Open string amplitudes on $AdS$ - building blocks}
\label{open}

We start with the $J_w(s,t)$ integrals 
\begin{eqnarray}
J_w(s,t) = \int_0^1 x^{s-1}(1-x)^{t-1} L_w(x) dx.
\end{eqnarray}
Here, $L_w(x)$ is the MPL labelled by the word $w$ in the alphabet with letters $\{0,1 \}$. See appendix \ref{conventions} for their definition and properties. The integral converges in the region $\Re(s)>0$, $\Re(t)>0$, but the analytic expressions we will find for $J_w(s,t)$ can be extended outside this region. The integrals can be computed in a low-energy expansion \cite{Alday:2024yax}
\begin{equation}
\label{Jlee}
J_w(s,t) = \text{poles} + \sum_{p,q=0} s^p t^q \sum_{W \in 0^p \shuffle 1^q \shuffle w}(L_{0W}(1)-L_{1W}(1)),
\end{equation}
but we are after analytic expressions valid for generic $s,t$. To proceed, it is convenient to introduce non-commutative variables $e_0,e_1$ - associated with the letters $0,1$ - and formally define the generating function
\begin{equation}\label{eq:GenSeriesOfMPLs}
L(e_0,e_1;x) = L_e(x) + L_0(x) e_0 +L_1(x) e_1 + L_{00}(x) e_0^2 +  L_{01}(x) e_0 e_1+ L_{10}(x) e_1 e_0 + \cdots.
\end{equation}
The derivative relations \eqref{eq:DerivativeOfMPLs} for MPLs are then equivalent to the Knizhnik–Zamolodchikov (KZ) equation
\begin{equation}
\label{LKZ}
\frac{\partial}{\partial x} L(e_0,e_1;x) = \left(\frac{e_0}{x}+ \frac{e_1}{x-1} \right)  L(e_0,e_1;x),
\end{equation}
with the boundary condition $L(e_0,e_1;x) = x^{e_0}$ as $x \to 0$. For later reference, it is convenient to introduce the Drinfeld associator 
\begin{equation}
\label{Drinfeld}
L(e_0,e_1;x) =L(e_1,e_0;1-x) Z(e_0,e_1),
\end{equation}
which satisfies the duality condition
\begin{equation}
Z(e_0,e_1)Z(e_1,e_0)=1.
\end{equation}
Correspondingly, we introduce a generating function for the $J_w(s,t)$ integrals 
\begin{align} \label{curlyJ}
\begin{split}
{\cal J}(s,t ; e_0,e_1) &=  \int_0^1 x^{s-1}(1-x)^{t-1} L(e_0,e_1;x)  dx \\ 
&= J_e(s,t)+ J_0(s,t) e_0 + J_1(s,t) e_1 + J_{00}(s,t) e_0^2+ J_{01}(s,t) e_0 e_1 +\cdots.
\end{split}
\end{align}
This generating function satisfies interesting properties. The shuffle identities \eqref{eq:ShuffleRelationsOfMPLs} of MPLs together with $\frac{\partial}{\partial s}x^s =L_0(x) x^s$ and $\frac{\partial}{\partial t}(1-x)^t =L_1(x) (1-x)^t$ imply
\begin{equation}
\partial_s J_w(s,t) = \sum_{w' \in 0 \shuffle w} J_{w'}(s,t),~~~\partial_t J_w(s,t) = \sum_{w' \in 1 \shuffle w} J_{w'}(s,t).
\end{equation}
At the level of the generating function, these imply
\begin{equation}
\label{derivativeJ}
\frac{\partial}{\partial s} {\cal J}(s,t; e_0,e_1) = \frac{\partial}{\partial e_0} {\cal J}(s,t; e_0,e_1),~~~\frac{\partial}{\partial t} {\cal J}(s,t; e_0,e_1) = \frac{\partial}{\partial e_1} {\cal J}(s,t; e_0,e_1)
\end{equation}
where the non-commutative derivative is defined such that 
\begin{equation}
\frac{\partial e_i}{\partial e_j}=\delta_{ij},
\end{equation}
together with the product rule
\begin{equation}
\frac{\partial}{\partial e_i} \left( f(e_0,e_1) g(e_0,e_1) \right)= \frac{\partial f(e_0,e_1)}{\partial e_i}  g(e_0,e_1)+ f(e_0,e_1) \frac{\partial g(e_0,e_1)}{\partial e_i} 
\end{equation}
where the order of products needs to be respected.\footnote{{\it e.g.} note that $\frac{\partial}{\partial e_0}(e_0 e_1 - e_1 e_0)=0$, so that $\frac{\partial}{\partial e_0}f(e_0,e_1)=0$ does not mean $f(e_0,e_1)$ is independent of $e_0$. } Furthermore, from its integral definition, it follows
\begin{eqnarray}
{\cal J}(s,t; e_0,e_1)={\cal J}(s+1,t; e_0,e_1)+{\cal J}(s,t+1; e_0,e_1).
\end{eqnarray}
The transformation property under shifts in only one of the two Mandelstam variables can be found by combining the KZ equation (\ref{LKZ}) with integration by parts. We find\footnote{Since $e_0$ and $e_1$ are non-commutative variables, expressions such as $\frac{1}{s+t+e_0+e_1}$ are to be interpreted as an expansion $\frac{1}{s+t+e_0+e_1}= \frac{1}{s+t} - \frac{e_0+e_1}{(s+t)^2}+ \cdots$.}
\begin{align}
\label{shiftJ}
\begin{split}
{\cal J}(s+1,t; e_0,e_1) &=\frac{1}{s+t+e_0+e_1} (s+e_0) {\cal J}(s,t; e_0,e_1) \hspace{2pt},\\
{\cal J}(s,t+1; e_0,e_1) &=\frac{1}{s+t+e_0+e_1} (t+e_1) {\cal J}(s,t; e_0,e_1). \end{split}
\end{align}
The shift relations imply a hierarchy for the functions $J_{w}(s,t)$ as the weight, namely the number of letters or the length of the word $w$, increases
\begin{align}
\begin{split}
(s+t){\cal J}(s+1,t; e_0,e_1) -s {\cal J}(s,t; e_0,e_1)= \frac{1}{s+t+e_0+e_1}(e_0\, t-e_1\, s) {\cal J}(s,t; e_0,e_1), \\
(s+t){\cal J}(s,t+1; e_0,e_1) -t {\cal J}(s,t; e_0,e_1)= \frac{1}{s+t+e_0+e_1}(e_1\, s-e_0\, t) {\cal J}(s,t; e_0,e_1) \hspace{2pt}.
\end{split}
\end{align}
These relations fix, in principle, $J_w(s,t)$ in terms of lower-weight integrals, modulo the Euler beta function, which is the solution of the homogeneous shift equations. The solution can be fixed by the behaviour as $s \to 0$. For $w \neq 0^p$, $J_w(s,t)$ is regular as $s \to 0$, while $J_{0^p}(s,t) = \frac{(-1)^p}{s^{p+1}} + \text{reg}$. In terms of generating functions\footnote{The limit $s \to 0$ is singular.}
\begin{eqnarray}
{\cal J}(s,t; e_0,e_1) = \frac{1}{s+e_0} + \text{reg},~~~\text{as $s \to 0$}.
\end{eqnarray}
MPLs $L_w(x)$ are closed under the argument transformation
\begin{equation}
x \to 1-x,
\end{equation} 
and their transformation properties are governed by the Drinfeld associator, see (\ref{Drinfeld}). Since the region of integration defining the $J_w(s,t)$ integrals  is invariant under such a transformation, the integrals have a symmetry under the exchange $s \leftrightarrow t$, also governed by the Drinfeld associator. In terms of the generating functions
\begin{equation}
\label{Jsym}
{\cal J}(t,s; e_1,e_0) = {\cal J}(s,t; e_0,e_1) Z(e_1,e_0).
\end{equation} 
This is consistent with the derivative relations (\ref{derivativeJ}) since
\begin{equation}
 \frac{\partial }{\partial e_0} Z(e_0,e_1) =  \frac{\partial }{\partial e_1} Z(e_0,e_1)=0.
\end{equation} 
Finally, let us mention that ${\cal J}(s,t; e_0,e_1) $ also satisfies a generalisation of Poincaré duality (\ref{Poincare}). This relation involves the $AdS$ Kernel, and is discussed in appendix \ref{poincare}. 

\subsection{Explicit results}
Up to weight four, the $J_w(s,t)$ integrals can be computed analytically, generally in terms of generalised hypergeometric functions. At weight zero, we have

\begin{equation}
J_e(s,t)= \beta(s,t)=\frac{\Gamma(s)\Gamma(t)}{\Gamma(s+t)}.
\end{equation}
The relations (\ref{derivativeJ}) then fix the integrals at weight one
\begin{equation}
J_0(s,t) = \partial_s J_e(s,t),~~~J_1(s,t) = \partial_t J_e(s,t).
\end{equation}
At weight two, three combinations are fixed by the derivative relations  (\ref{derivativeJ})
\begin{equation}
J_{00}(s,t) = \frac{1}{2}\partial^2_s J_e(s,t),~~~J_{11}(s,t) = \frac{1}{2}\partial^2_t J_e(s,t),~~~J_{01}(s,t)+ J_{10}(s,t) = \partial_s \partial_t J_e(s,t) \hspace{2pt}.
\end{equation}
In addition, we have 
\begin{equation}
J_{01}(s,t)=-\frac{\Gamma(1+s)\Gamma(t)}{\Gamma(1+s+t)} \setlength\arraycolsep{1pt}
{}_{4} F_{3}\left(\begin{matrix}1,& 1,& 1,& &1+s&\\
2,&
 & 2,& &1+s+t&  \end{matrix};1\right) \hspace{2pt},
\end{equation}
which can be computed by direct integration. Here we have used the general result
\begin{equation}
L_{0^{n-1}1}(x)= -\text{Li}_n(x).
\end{equation}
At weight three, there are eight independent integrals. Two of them are given by
\begin{equation}
J_{001}(s,t)=-\frac{\Gamma(1+s)\Gamma(t)}{\Gamma(1+s+t)} \setlength\arraycolsep{1pt}
{}_{5} F_{4}\left(\begin{matrix}1,& 1,& 1,&1,& &1+s&\\
2,&
 & 2,& & 2, &1+s+t&  \end{matrix};1\right),
\end{equation}
and
\begin{equation}
J_{110}(s,t)=J_{001}(t,s)+\zeta(2) J_0(t,s)+\zeta(3)J_e(t,s).
\end{equation}
The other six combinations can be written in terms of derivatives
\begin{eqnarray}
&J_{000}(s,t) = \frac{1}{6}\partial^3_s J_e(s,t),~~~J_{111}(s,t) = \frac{1}{6}\partial^3_t J_e(s,t),\nonumber\\
&\partial_s J_{11}(s,t) =J_{011}(s,t) +J_{101}(s,t) +J_{110}(s,t),~~~ \partial_t J_{00}(s,t) =J_{100}(s,t) +J_{010}(s,t) +J_{001}(s,t), \nonumber \\
&\partial_s J_{01}(s,t)=2J_{001}(s,t)+J_{010}(s,t),~~~\partial_t J_{10}(s,t)=2J_{110}(s,t)+J_{101}(s,t). 
\end{eqnarray}
As we increase the weight, the integrals become more and more complicated. We will show below that they are particular cases of Aomoto-Gelfand generalised hypergeometric functions. At weight four, they can still be written in terms of hypergeometric functions and their derivatives. Let us start with words with two zeroes and two ones, which are the most complicated examples. There are six of them. The following combinations can be written in terms of derivatives of lower-order integrals, already computed
\begin{align}
\begin{split}
\partial_s J_{011}(s,t) &=2J_{0011}(s,t)+J_{0101}(s,t)+J_{0110}(s,t) \\
\partial_s J_{101}(s,t) &=J_{0101}(s,t)+2J_{1001}(s,t)+J_{1010}(s,t) \\
\partial_s J_{110}(s,t) &= J_{0110}(s,t)+J_{1010}(s,t)+2J_{1100}(s,t) \\
\partial_t J_{001}(s,t) &= J_{1001}(s,t)+J_{0101}(s,t)+2J_{0011}(s,t) \\
\partial_t J_{010}(s,t) &= J_{1010}(s,t)+2J_{0110}(s,t)+J_{0101}(s,t) \\
\partial_t J_{100}(s,t) &= 2J_{1100}(s,t)+J_{1010}(s,t)+J_{1001}(s,t) \hspace{2pt}.
\end{split}
\end{align}
These relations are not all independent, and fix only four integrals. Two extra integrals can be computed as follows. Let us start with $J_{0011}(s,t)$. Plugging the series representation around zero for $L_{0011}(x)$ (see appendix \ref{conventions}) and integrating term by term, we obtain
\begin{equation}
J_{0011}(s,t)= \sum_{\ell=1}^\infty \frac{\Gamma(s+\ell)\Gamma(t)}{\Gamma(\ell+s+t)} \frac{H(\ell-1)}{\ell^3},
\end{equation}
where $H(\ell-1)$ is the harmonic number. To perform the sum, we notice that
\begin{equation}
H(\ell-1)=\gamma_e+ \left. \partial_\epsilon \frac{\Gamma(\ell+\epsilon)}{\Gamma(\ell)} \right|_{\epsilon=0} \hspace{2pt}.
\end{equation}
The sum over $\ell$ can now be performed and we obtain
\begin{equation}
J_{0011}(s,t)=\frac{\Gamma(1+s)\Gamma(t)}{\Gamma(1+s+t)} \setlength\arraycolsep{1pt}
\partial_\epsilon \left. {}_{5} F_{4}\left(\begin{matrix}1,& 1,& 1,&1+\epsilon,& &1+s&\\
2,&
 & 2,& & 2, &1+s+t&  \end{matrix};1\right) \right|_{\epsilon=0} \hspace{2pt}.
\end{equation}
Using similar tricks, we can compute all other integrals, for instance
\begin{equation}
J_{0101}(s,t)=\frac{\Gamma(1+s)\Gamma(t)}{2\Gamma(1+s+t)} \setlength\arraycolsep{1pt}
\left(\partial_{\epsilon_1}^2-\partial_{\epsilon_2}^2 \right) \left. {}_{5} F_{4}\left(\begin{matrix}1,& 1,& 1,&1+\epsilon_2,& &1+s&\\
2,&
 & 2,& & 1+\epsilon_1, &1+s+t&  \end{matrix};1\right) \right|_{\epsilon_1=\epsilon_2=0} \hspace{2pt}.
\end{equation}
Together with the relations given above, this fixes all integrals at weight four with two zeroes and two ones. Let us now focus on the integrals with three zeroes and one one. Integrals with three ones and one zero are related to those by $s \leftrightarrow t$ symmetry (\ref{Jsym}). By direct integration, or by applying the trick above we obtain
\begin{equation}
J_{0001}(s,t) =- \frac{\Gamma(1+s)\Gamma(t)}{\Gamma(1+s+t)} \setlength\arraycolsep{1pt}
{}_{6} F_{5}\left(\begin{matrix}1,& &\cdots & &,1,& &1+s\\
2,&
&\cdots& & ,2,& &1+s+t & \end{matrix};1\right).
\end{equation}
The other three independent combinations can be fixed by the derivative relations, as can $J_{0000}(s,t)$ and $J_{1111}(s,t)$. 

Before proceeding, let us mention that the $J_w(s,t)$ integrals have an interesting structure of poles. This is worked out in  appendix \ref{poles}.\footnote{This is useful for applications to dispersive sum rules and in order to compute the spectrum of intermediate operators from the amplitude.}

\subsection{Relation to Aomoto-Gelfand hypergeometric functions}
We can always use the shuffle relations to write multiple polylogarithms in terms of $L_0(x)$ and multiple polylogarithms whose label ends in the letter $1$. For instance,
\begin{equation}
L_{10}(x) = -L_{01}(x) +L_{1}(x) L_{0}(x).
\end{equation}
This implies that all $J_w(s,t)$ integrals can be written in terms of the ones whose label $w$ ends with the letter 1, plus derivatives w.r.t. $s$. Without loss of generality, let us then focus on words ending in the letter 1. The representation of MPLs in terms of iterated integrals then leads to  
\begin{equation}
J_{a_1a_2 \cdots a_{r-1}1}(s,t) = \int_0^1 dx x^{s-1} (1-x)^{t-1} \int_0^x \frac{dx_1}{x_1-a_1} \int_0^{x_1} \frac{dx_2}{x_2-a_2} \cdots \int_0^{x_{r-1}} \frac{dx_r}{x_r-1}.
\end{equation}
We can now make a change of coordinates 
\begin{equation}
u_0=x,~~u_n=\frac{x_n}{x_{n-1}},~~~\text{for $n=1,2,\cdots$}
\end{equation}
so that $u_n \in [0,1]$. The integral then becomes 
\begin{equation}
J_{a_1a_2\cdots a_{r-1}1}(s,t) =\int_0^1 \prod_{i=0}^r du_i u_0^{s-1}(1-u_0)^{t-1} \frac{u_0^r u_1^{r-1} u_2^{r-2} \cdots u^2_{r-2} u_{r-1}}{(u_0 u_1-a_1)(u_0 u_1 u_2 - a_2) \cdots (u_0 \cdots  u_r-1)}.
\label{Jintegralrep}
\end{equation}
This is a particular case of Aomoto-Gelfand hypergeometric function of type ${\left(r+2,2r+d+2\right)}$, see \cite{Gel86,Aomoto1974,Duhr:2023bku}
\begin{equation}
\int_0^1 \prod_{i=0}^r du_i u_i^{\alpha_i-1}(1-u_i)^{\beta_i-\alpha_i-1}(1- y_r u_0 \cdots  u_r)^{-\gamma_r} \cdots (1- y_1 u_0 u_1)^{-\gamma_1},
\end{equation}
where to make contact with (\ref{Jintegralrep}) we choose $\gamma_i \neq 0$ only for $a_i=1$. Here $r$ is the weight, or length of the word, and $d$ is the depth, or number of ones. The integrals we consider in this paper are obtained in the limit $y_i \to 1$. Let us consider for example $J_{0^{r-1}1}(s,t)$. In this case
\begin{equation}
J_{0^{r-1}1}(s,t) = J_{0^{r-1}1}(s,t;1)
\end{equation}
where
\begin{equation}
J_{0^{r-1}1}(s,t;y) =-\int_0^1 \prod_{i=0}^r du_i \frac{u_0^s(1-u_0)^{t-1}}{1-y u_0 u_1 \cdots u_r} = -\frac{\Gamma(1+s)\Gamma(t)}{\Gamma(1+s+t)} \setlength\arraycolsep{1pt}
{}_{r+2} F_{r+1}\left(\begin{matrix}1,& &\cdots & &,1,& &1+s\\
2,&
&\cdots& & ,2,& &1+s+t & \end{matrix};y\right)
\end{equation}
is a standard generalised hypergeometric function. The relation to Aomoto-Gelfand hypergeometric functions suggests the introduction of extra variables $\mathrm{y}_i$ such that at $\mathrm{y}_i=1$, the integrals reduce to the expressions that appear in the open string scattering problem. It is convenient to do this in the compact notation where
\begin{equation}
J_{s_1,s_2,\cdots ,s_d}(s,t)= (-1)^d J_{0^{s_1-1} 1 0^{s_2-1} \cdots 1}(s,t),~~~s_i \geq 1 \hspace{2pt}.
\end{equation}
By writing $L_{s_1,\cdots,s_d}(x)$ as a nested sum and integrating term by term, we obtain
\begin{equation}
J_{s_1,\cdots,s_d}(s,t) = \sum^\infty_{\ell_1>\ell_2>\cdots>\ell_d>0} \frac{\beta(s+\ell_1,t)}{\ell_1^{s_1}\ell_2^{s_2} \cdots \ell_d^{s_d}} \hspace{2pt},
\end{equation}
where $\beta(s+\ell,t)= \frac{\Gamma(s +\ell)\Gamma(t)}{\Gamma(s+t+\ell)}$. We can now introduce the extra variables $\mathrm{y}_i$
\begin{equation}
J_{s_1,\cdots,s_d}(s,t;\mathrm{y}_1,\cdots,\mathrm{y}_d) = \sum^\infty_{\ell_1>\ell_2>\cdots>\ell_d>0} \frac{\beta(s+\ell_1,t) \mathrm{y}_1^{\ell_1} \cdots \mathrm{y}_d^{\ell_d}}{\ell_1^{s_1}\ell_2^{s_2} \cdots \ell_d^{s_d}}
\end{equation}
so that 
\begin{equation}
J_{s_1,s_2,\cdots,s_d}(s,t) =J_{s_1,s_2,\cdots,s_d}(s,t;1,\cdots,1). 
\end{equation}
The functions $J_{s_1,s_2,\cdots,s_d}(s,t;\mathrm{y}_1,\cdots,\mathrm{y}_d)$ are generalised hypergeometric functions.\footnote{More precisely, they fall into the category of Horn's hypergeometric series, in general given by $$\sum_{n_1,n_2,\cdots=0} c(n_1,n_2,\cdots) z_1^{n_1} z_2^{n_2} \cdots$$ with $c(n_1+1,n_2,\cdots)/c(n_1,n_2,\cdots),c(n_1,n_2+1,\cdots)/c(n_1,n_2,\cdots)$ and so on, rational functions of the $n_i$.}  As such, they satisfy differential relations in the variables $\mathrm{y}_i$. Indeed
\begin{eqnarray}
\mathrm{y}_q \frac{\partial}{\partial \mathrm{y}_q} J_{s_1,\cdots,s_d}(s,t;\mathrm{y}_1,\cdots,\mathrm{y}_d) =J_{s_1,\cdots, s_q-1,\cdots,s_d}(s,t;\mathrm{y}_1,\cdots,\mathrm{y}_d)
\end{eqnarray}
so that the operator $\mathrm{y}_q \frac{\partial}{\partial \mathrm{y}_q}$ lowers the index $s_q$, and the total weight, by one. By repeated action of such operators, we reach
\begin{equation}
J_{0,\cdots,0}(s,t;\mathrm{y}_1,\cdots,\mathrm{y}_d) \equiv \sum^\infty_{\ell_1>\ell_2>\cdots>\ell_d>0} \beta(s+\ell_1,t) \mathrm{y}_1^{\ell_1} \cdots \mathrm{y}_d^{\ell_d} \hspace{2pt}.
\end{equation}
These are particular cases of Lauricella hypergeometric functions, and for the present case, they can be written as linear combinations of hypergeometric functions. 

\section{Closed string amplitudes on $AdS$ - building blocks}
\label{closed}
We now turn our attention to the building blocks of closed string amplitudes on $AdS$

\begin{equation}
\label{Iint}
I_w(s,t) =  \int |z|^{2s-2}|1-z|^{2t-2} {\cal L}_w(z) d^2 z 
\end{equation}
where ${\cal L}_w(z)$ is the single-valued multiple polylogarithm (SVMPL) labelled by the word $w$ in the alphabet with letters $\{0,1 \}$. The integral converges for ${\Re(s)>0,\Re(t)>0},{\Re(s+t)<1}$, but the explicit results we will find can be continued beyond that region. The integrals can be computed in a low-energy expansion \cite{Alday:2023jdk}
\begin{equation}
\label{Ilee}
I_w(s,t) = \text{poles} + \sum_{p,q=0} s^p t^q \sum_{W \in 0^p \shuffle 1^q \shuffle w}({\cal L}_{0W}(1)-{\cal L}_{1W}(1)).
\end{equation}
From (\ref{Jlee}) and (\ref{Ilee}) we can see that 
\begin{equation}
I_w(s,t)=sv\left(J_w(s,t)\right) \hspace{2pt},
\end{equation}
since ${\cal L}_{w}(1)=sv\left(L_{w}(1)\right)$, so that the closed string amplitude building blocks are the single-valued version of the open string amplitude building blocks, as expected. We are, however, after a quadratic relation {\`a} la KLT (\ref{KLT}). As in the previous section, it is convenient to introduce a generating function for the SVMPLs
\begin{equation}
{\cal L}(e_0,e_1;z) = {\cal L}_e(z)+ {\cal L}_0(z)e_0 + {\cal L}_1(z)e_1 +{\cal L}_{00}(z)e_0^2 + {\cal L}_{01}(z)e_0 e_1+ {\cal L}_{10}(z)e_1 e_0+ \cdots,
\end{equation}
constructed from the generating series of MPLs \eqref{eq:GenSeriesOfMPLs} according to \eqref{eq:ConstructionOfSVMPLsFromMPLs}. It is a formal solution of holomorphic and anti-holomorphic KZ equations
\begin{align}
\begin{split}
\frac{\partial}{\partial z} {\cal L}(e_0,e_1;z)  &= \left(\frac{e_0}{z}+\frac{e_1}{z-1} \right){\cal L}(e_0,e_1;z) \\
\frac{\partial}{\partial \bar z} {\cal L}(e_0,e_1;z) &= {\cal L}(e_0,e_1;z)  \left(\frac{e_0}{\bar z}+\frac{e'_1}{\bar z-1} \right),
\end{split}
\end{align}
where the deformed variable $e'_1$ is given in appendix \ref{conventions}, and we have the boundary conditions $ {\cal L}(e_0,e_1;z) = e^{e_0 \log|z|^2}$ as $z \to 0$. One can introduce the Deligne associator as 
\begin{equation}
{\cal L}(e_0,e_1;z)={\cal L}(e_1,e_0;1-z)W(e_0,e_1),
\end{equation}
which satisfies the duality condition $W(e_0,e_1)W(e_1,e_0)=1$. Integrating the generating function ${\cal L}(e_0,e_1;z)$ term by term leads to the generating function for the  $I_w(s,t)$ integrals:
\begin{equation}
{\cal I}(s,t;e_0,e_1) = \int |z|^{2s-2}|1-z|^{2t-2} {\cal L}(e_0,e_1;z) d^2 z.
\end{equation}
We will start by discussing the general properties of ${\cal I}(s,t;e_0,e_1)$. Then, we will turn to its explicit computation and relationship to the generating function ${\cal J}(s,t;e_0,e_1)$ appearing in the problem of open-string scattering on $AdS$. The shuffle identities together with $\frac{\partial}{\partial s}|z|^{2s} = {\cal L}_0(z) |z|^{2s}$ and $\frac{\partial}{\partial t}|1-z|^{2t} = {\cal L}_1(z) |1-z|^{2t}$ imply

\begin{equation}
\label{derivativeI}
\frac{\partial}{\partial s} {\cal I}(s,t; e_0,e_1) = \frac{\partial}{\partial e_0} {\cal I}(s,t; e_0,e_1),~~~\frac{\partial}{\partial t} {\cal I}(s,t; e_0,e_1) = \frac{\partial}{\partial e_1} {\cal I}(s,t; e_0,e_1) \hspace{2pt}.
\end{equation}
Furthermore, the holomorphic and anti-holomorphic KZ equations for $ {\cal L}(e_0,e_1;z)$ together with integration by parts lead to
\begin{align}
\label{shiftI}
\begin{split}
{\cal I}(s+1,t; e_0,e_1)= \frac{1}{s+t+e_0+e_1}(s+e_0) {\cal I}(s,t; e_0,e_1) (s+e_0) \frac{1}{s+t+e_0+e'_1} \hspace{2pt},\\
{\cal I}(s,t+1; e_0,e_1)= \frac{1}{s+t+e_0+e_1}(t+e_1) {\cal I}(s,t; e_0,e_1) (t+e'_1) \frac{1}{s+t+e_0+e'_1} \hspace{2pt}.
\end{split}
\end{align}
In order to obtain these relations, we have done formal manipulations that take us outside the convergence region of the relevant integrals. We assume that these relations hold for the appropriate analytic continuation. SVMPLs are closed under $z \to 1-z$, and their transformation is governed by the Deligne associator. Performing a change of coordinates $z \to 1-z$ in (\ref{Iint}) implies
\begin{equation}
{\cal I}(t,s; e_1,e_0) = {\cal I}(s,t; e_0,e_1) W(e_1,e_0).
\end{equation} 
The properties obtained so far mimic the properties for the building blocks of open string amplitudes in $AdS$. In addition, it follows from (\ref{Iint}) that the result for $I_w(s,t)$ is only sensitive to the symmetric part of ${\cal L}_w(z)$ under the exchange of $z \leftrightarrow \bar z$. Together with the explicit construction of SVMPLs, see appendix \ref{conventions}, this leads to the following relation
\begin{equation}
\label{reality}
{\cal I}(s,t; e_0,e_1) = {\cal I}^R(s,t; e_0,e'_1)
\end{equation} 
where 
\begin{equation}
{\cal I}^R(s,t; e_0,e'_1) = I_e(s,t) + \cdots + I_{00010} e_0 e'_1 e_0 e_0 e_0 + \cdots
\end{equation} 
i.e. the order of the letters in the word labelling $I_w(s,t)$ is reversed. 

\subsection{Explicit results}
We will now compute the generating function ${\cal I}(s,t; e_0,e_1)$ in a series expansion in the non-commutative variables $e_0,e_1$:
\begin{equation}
{\cal I}(s,t; e_0,e_1)  = I_e(s,t)+\cdots
\end{equation}
with $I_e(s,t)=\beta_{\mathbb{C}}(s,t)$ the complex beta function. The condition (\ref{reality}) imposes relations among different components. Up to weight four
\begin{eqnarray} 
&I_{10}(s,t) = I_{01}(s,t),~~~I_{100}(s,t) = I_{001}(s,t),~~~I_{110}(s,t) = I_{011}(s,t)\nonumber\\
&I_{1110}(s,t)=I_{0111}(s,t)+2\zeta(3)I_1(s,t),~~~I_{1100}(s,t)=I_{0011}(s,t)-2\zeta(3)I_1(s,t)\nonumber\\
&I_{1101}(s,t)=I_{1011}(s,t)-6\zeta(3)I_1(s,t),~~~I_{1010}(s,t)=I_{0101}(s,t)+4\zeta(3)I_1(s,t)\nonumber\\
&I_{1000}(s,t)=I_{0001}(s,t),~~~I_{0100}(s,t)=I_{0010}(s,t) \hspace{2pt},
\end{eqnarray}
leaving three independent integrals at weight two, six at weight three and ten at weight four. Next, we impose the derivative relations (\ref{derivativeI}). Up to weight two, this fixes the remaining integrals, in terms of the complex beta function, equivalently $I_e(s,t)$:
\begin{align}
\begin{split}
& \hspace{80pt} I_{0}(s,t) = \partial_s I_e(s,t),~~~I_{1}(s,t) = \partial_t I_e(s,t),\\
&I_{00}(s,t) = \frac{1}{2} \partial^2_s I_e(s,t),~~~I_{11}(s,t) = \frac{1}{2} \partial^2_t I_e(s,t),~~~I_{01}(s,t) = \frac{1}{2} \partial_s \partial_t I_e(s,t) \hspace{2pt}.
\end{split}
\end{align}
At weight three, it leaves only two independent integrals, which we take to be $I_{001}(s,t)$ and $I_{011}(s,t)$:
\begin{align}
\begin{split}
& \hspace{85pt} I_{000}(s,t) = \frac{1}{6} \partial^3_s I_e(s,t),~~~I_{111}(s,t) = \frac{1}{6} \partial^3_t I_e(s,t),\\
&I_{101}(s,t) =-2 I_{011}(s,t) +\frac{1}{2} \partial_s \partial_t^2 I_e(s,t),~~~I_{010}(s,t) =-2 I_{001}(s,t) +\frac{1}{2} \partial^2_s \partial_t I_e(s,t) \hspace{2pt}.
\end{split}
\end{align}
At weight four, it turns out the relations (\ref{derivativeI}) are quite powerful, and only one independent integral remains, which we take to be $I_{0101}(s,t)$
\begin{eqnarray}
I_{0000}(s,t) &=& \frac{1}{24} \partial^4_s I_e(s,t),~~~I_{1111}(s,t) = \frac{1}{24} \partial^4_t I_e(s,t),\nonumber\\
I_{1011}(s,t) &=& -\frac{1}{2} \partial_t I_{011}(s,t)+\frac{1}{8} \partial_s \partial_t^3 I_e(s,t)+3 \zeta(3)\partial_t I_e(s,t)\nonumber\\
I_{0010}(s,t) &=& -\frac{1}{2} \partial_s I_{001}(s,t)+\frac{1}{8} \partial^3_s \partial_t I_e(s,t)\nonumber\\
I_{0001}(s,t) &=& \frac{1}{2} \partial_s I_{001}(s,t)-\frac{1}{24} \partial^3_s \partial_t I_e(s,t)\\
I_{0110}(s,t) &=& -I_{0101}(s,t) - \partial_t I_{001}(s,t)+\frac{1}{4} \partial^2_s \partial^2_t I_e(s,t)-2\zeta(3) \partial_t I_e(s,t)\nonumber\\
I_{1001}(s,t) &=& -I_{0101}(s,t) - \partial_s I_{011}(s,t)+\frac{1}{4} \partial^2_s \partial^2_t I_e(s,t)-2\zeta(3) \partial_t I_e(s,t)\nonumber\\
I_{0011}(s,t) &=& \frac{1}{2} \partial_s I_{011}(s,t) + \frac{1}{2}\partial_t  I_{001}(s,t)  -\frac{1}{8} \partial^2_s \partial^2_t I_e(s,t)+\zeta(3) \partial_t I_e(s,t)\nonumber\\
I_{0111}(s,t) &=& \frac{1}{2}\partial_t  I_{011}(s,t)  -\frac{1}{24} \partial_s \partial^3_t I_e(s,t)-\zeta(3) \partial_t I_e(s,t).\nonumber
\end{eqnarray}
Our next task is to compute explicitly the independent integrals. We do so by extending the ideas of KLT  \cite{Kawai:1985xq} to the integration of  SVMPLs, following \cite{Alday:2024yax}. In the appendix to that paper, we considered integrals of the form
\begin{equation}
A_{\rm{closed}}(s,t)= \int |z|^{2s-2}|1-z|^{2t-2} \sum_i^n F_i(z) G_i(\bar z) d^2 z,
\end{equation}
where the inserted sum is finite and single-valued, but each independent term is not necessarily single-valued. By deforming the contours as in \cite{Kawai:1985xq}, it was proven that these integrals can be factorised into one-dimensional integrals
\begin{eqnarray}
\label{holofac}
A_{\rm{closed}}(s,t)= \frac{1}{2\pi i}\sum_i^n \int_0^1\!\! x^{s-1}(1-x)^{t-1} G_i(x) dx \! \int_1^\infty \!\! \!\! y^{s-1} \text{Disc}_1\left[(1-y)^{t-1}F_i(y) \right] dy,
\end{eqnarray}
where the discontinuity across the real axis for $y>1$ is defined by
\begin{equation}
 \text{Disc}_1\left[f(y) \right] = f(y+i \epsilon) -f(y-i \epsilon),~~~y>1.
\end{equation}
For $\sum_i^n F_i(z) G_i(\bar z)=1$, we get $ \text{Disc}_1\left[(1-y)^{t-1} \right] =2 i \sin(\pi t)(y-1)^{t-1}$ and
\begin{equation}
A_{\rm{closed}}(s,t)=\frac{1}{\pi} \sin(\pi t) \int_0^1 x^{s-1}(1-x)^{t-1}dx \int_1^\infty y^{s-1}(y-1)^{t-1} dy,
\end{equation}
which is the KLT formula. Given the insertion of SVMPLs ${\cal L}_w(z)$, we can always factorise the integrals into products of 1d integrals involving MPLs $L_w(x)$. These are almost identical to the integrals considered in the previous section, except the second integral is in the range $y \in [1,\infty)$ and one needs to keep track of discontinuities. The resulting expressions are very lengthy, but the result can always be written as bilinears of the 1d  $J_w(s,t)$ integrals. For example
\begin{eqnarray}
I_{001}(s,t) &=& \kappa(s,t)\left( J_{001}(s,t) J_e(s,t) + J_{100}(s,t) J_e(s,t) +J_{00}(s,t)J_1(s,t)+J_{0}(s,t)J_{10}(s,t) \right) \nonumber \\
&&+  \partial_t \kappa(s,t)  J_{00}(s,t) J_e(s,t)+ \partial_s  \kappa(s,t) (J_{10}(s,t) J_e(s,t)+J_{0}(s,t) J_1(s,t))\nonumber\\
&&+\frac{1}{2} \partial_s \partial_t \kappa(s,t)  J_{0}(s,t) J_e(s,t)+\frac{1}{2} \partial^2_s \kappa(s,t)  J_{1}(s,t) J_e(s,t) \\
&&+ \pi^2 \frac{2 \cos ^2(\pi  (s+t))-\cos (2 \pi  s)-\cos (2 \pi  t)}{4 \sin(\pi(s+t))^4}J_e(s,t)J_e(s,t) \nonumber,
\end{eqnarray}
where $\kappa(s,t)= \frac{\sin(\pi s) \sin(\pi t)}{\pi \sin(\pi(s+t))}$ is the KLT Kernel. This suggests writing our results in the following illuminating way.

\subsection{$AdS$ KLT relations}

Motivated by the KLT relations, the construction of SVMPLs from MPLs, and the result above, we propose the following relation between the building blocks of open and closed string amplitudes on $AdS$
\begin{equation}
\label{adskernel}
 {\cal I}(s,t; e_0,e_1) =  {\cal J}(s,t; e_0,e_1) {\cal K}(s,t;e_0,e_1)  {\cal J}^R(s,t; e_0,e'_1),
\end{equation}
where ${\cal K}(s,t;e_0,e_1)$ admits an expansion in the non-commutative variables
\begin{equation}
{\cal K}(s,t;e_0,e_1) =\kappa(s,t) + \kappa_0(s,t) e_0 + \kappa_1(s,t) e_1 + \cdots,
\end{equation}
starts with the KLT Kernel $\kappa(s,t)$ and contains only trigonometric functions in $s,t$. Indeed, as a consequence of the shift relations (\ref{shiftJ}) and (\ref{shiftI}), the AdS KLT Kernel is periodic in both $s$ and $t$
\begin{equation}
{\cal K}(s+1,t;e_0,e_1) ={\cal K}(s,t+1;e_0,e_1) ={\cal K}(s,t;e_0,e_1).
\end{equation}
As a consequence of the derivative relations (\ref{derivativeJ}) and (\ref{derivativeI}), the Kernel also satisfies
\begin{equation}
\label{derivativeK}
\partial_s {\cal K}(s,t;e_0,e_1) = \frac{\partial}{\partial e_0} {\cal K}(s,t;e_0,e_1),~~~\partial_t {\cal K}(s,t;e_0,e_1) = \frac{\partial}{\partial e_1} {\cal K}(s,t;e_0,e_1),
\end{equation}
where we have used the fact that  (recall $\frac{\partial}{\partial e_1}$ is a non-commutative derivative)
\begin{equation}
\frac{\partial}{\partial e_1} e'_1=1.
\end{equation}
In addition, (\ref{reality}) leads to 
\begin{equation}
\label{realityK}
 {\cal K}(s,t;e_0,e_1) = {\cal K}^R(s,t;e_0,e'_1).
\end{equation}
Finally, the $AdS$ Kernel also possesses a symmetry under the exchange of $s$ and $t$, inherited from the respective symmetries for the 1d and 2d integrals
\begin{equation}
\label{Ksym}
 {\cal K}(t,s;e_1,e_0) = Z(e_0,e_1){\cal K}(s,t;e_0,e_1) Z(e_1,e_0).
\end{equation}
To a given order, one can compute the AdS Kernel by computing the generating functions ${\cal J}(s,t;e_0,e_1)$ and ${\cal I}(s,t;e_0,e_1)$ explicitly. The relation (\ref{adskernel}) hence has a unique solution for ${\cal K}(s,t;e_0,e_1)$, which can be computed order by order in $e_0,e_1$. The following way, however, is much more instructive and allows us to compute the AdS KLT Kernel to all orders. Consider the generating function
\begin{align}
\begin{split}
{\cal I}(s,t;e_0,e_1) &= \int |z|^{2s-2}|1-z|^{2t-2} {\cal L}(e_0,e_1;z) d^2 z \\
&= \int |z|^{2s-2}|1-z|^{2t-2}L(e_0,e_1;z) L^R(e_0,e'_1;\bar z) d^2 z.
\end{split}
\end{align}
The holomorphic factorisation formula (\ref{holofac}) then leads to
\begin{equation}
{\cal I}(s,t;e_0,e_1) = \frac{1}{2\pi i} \int_1^\infty dy y^{s-1}\text{Disc}_1 \left[(1-y)^{t-1} L(e_0,e_1;y) \right] \int_0^1 dx x^{s-1}(1-x)^{t-1} L^R(e_0,e'_1;x),
\end{equation}\\[-10pt]
where the order of the two factors is important and recall $\text{Disc}_1[f(y)]=f(y+i\epsilon)-f(y-i \epsilon)$ for $y>1$. Note that the second factor is nothing but the generating functional ${\cal J}^R(s,t;e_0,e'_1)$. Let us now focus on the first factor. In order to convert the integral to an integral in the correct range $[1,\infty) \to [0,1]$, consider the following identities
\begin{equation}
\label{contourdef}
\left.\left(\int_{-\infty}^0 dz + \int_0^1 dz + \int_1^\infty dz \right) z^{s-1}(1-z)^{t-1}F(z) \right|_{z=x \pm i \epsilon}=0,
\end{equation}
where in our case $F(z) =L(e_0,e_1;z)$ and the contour in each case is just above or just below the real line. Furthermore, we have
\begin{align}
\begin{split}
\left. z^{s-1} \right|_{z=x\pm i \epsilon} &= -e^{\pm i \pi s}(-x)^{s-1} ,~~~x<0, \\
\left. (1-z)^{t-1} \right|_{z=x \pm i\epsilon} &= -e^{\mp i \pi t} (x-1)^{t-1},~~~x>1.
\end{split}
\end{align}
 The discontinuities of MPLs as we cross the real line, either for $x<0$ or $x>1$, are governed by monodromy matrices:
\begin{align}
\begin{split}
 L(e_0,e_1; x + i \epsilon) &= L(e_0,e_1; x - i \epsilon) M_0,~~~x<0 \\
 L(e_0,e_1; x - i \epsilon) &= L(e_0,e_1; x + i \epsilon) M_1,~~~x>1,
\end{split}
\end{align}
where the monodromy matrices have been computed in \cite{FrancisB}  and are given by 
\begin{equation}
M_0=e^{2\pi i e_0},~~~M_1=Z(e_1,e_0 )e^{2\pi i e_1} Z(e_0,e_1).
\end{equation} 
For clarity of notation, let us introduce the following three objects
\begin{eqnarray}
J^{(-)}&=& \int_{-\infty}^0 dx (-x)^{s-1} (1-x)^{t-1}  L(e_0,e_1; x-i \epsilon),\nonumber\\
J&=& \int_0^1 dx x^{s-1} (1-x)^{t-1}  L(e_0,e_1; x),\\
J^{(+)}&=& \int_1^\infty dx x^{s-1} (x-1)^{t-1}  L(e_0,e_1; x+ i \epsilon),\nonumber
\end{eqnarray}
where we have suppressed the $(s,t;e_0,e_1)$ dependence. Notice that $J$ in (3.36) is the same as $\mathcal{J}$ in \eqref{curlyJ}. As a consequence of (\ref{contourdef}), the three objects satisfy
\begin{align}
\begin{split}
-e^{i \pi s} J^{(-)}M_0+ J- e^{-i \pi t}J^{(+)} &=0 \\
-e^{-i \pi s} J^{(-)}+ J- e^{i \pi t}J^{(+)}M_1 &=0,
\end{split}
\end{align}
where the monodromy matrices do not commute with $J,J^{(-)},J^{(+)}$. We can always write
\begin{eqnarray} \label{defK}
e^{-i \pi t}J^{(+)}-e^{i \pi t}J^{(+)}M_1 = J K,
\end{eqnarray}
where $K$ is invertible\footnote{A function $f(e_0,e_1)=\alpha + \alpha_0 e_0 +\alpha_1 e_1+\cdots$ with $\alpha \neq 0$ is always invertible. In particular, $J$, $J^{(-)}$, and $J^{(+)} $ are invertible, hence the validity of \eqref{defK}.} and related to the $AdS$ KLT Kernel by ${\cal K}=-\frac{1}{2\pi i} K$. Plugging this in the previous two relations we find
\begin{align}
\begin{split}
-e^{i \pi s} J^{(-)}M_0+ J^{(+)}\left( -e^{-i \pi t}+ e^{-i \pi t} K^{-1}-e^{i \pi t} M_1 K^{-1} \right) &=0,\\
-e^{-i \pi s} J^{(-)}+J^{(+)}\left( - e^{i \pi t } M_1+ e^{-i \pi t} K^{-1}-e^{i \pi t} M_1 K^{-1}  \right) &=0.
\end{split}
\end{align}
Since $J^{(-)}$ and $J^{(+)}$ are invertible, $K^{-1}$ should satisfy
\begin{eqnarray*}
e^{-i \pi s} \left( -e^{-i \pi t}+ e^{-i \pi t} K^{-1}-e^{i \pi t} M_1 K^{-1} \right)=e^{i \pi s} \left( - e^{i \pi t}M_1+ e^{-i \pi t} K^{-1}-e^{i \pi t} M_1 K^{-1}  \right)M_0.
\end{eqnarray*}
We can solve for $K^{-1}$ and find
\begin{eqnarray}
\label{Kinverse}
K^{-1}=1+\frac{e^{2 i \pi s} M_0}{1-e^{2 i \pi s} M_0}+\frac{e^{2 i \pi t} M_1}{1-e^{2 i \pi t} M_1}.
\end{eqnarray}
At zeroth order in the non-commutative variables, $M_0=M_1=1$ and we get the inverse of the usual KLT Kernel
\begin{equation}
{\cal K}^{-1} = -2\pi i K^{-1}= \pi  (\cot (\pi  s)+\cot (\pi  t))+ \cdots .
\end{equation}
At higher orders, and up to weight four, it precisely relates the explicit integrals we computed. Plugging the explicit expressions for the monodromy matrices, we find
\begin{equation}
{\cal K}^{-1} = -2\pi i K^{-1}= \pi  \left(\cot \left( \pi  (s+e_0) \right)+Z(e_1,e_0)\cot \left( \pi  (t+e_1) \right)Z(e_0,e_1) \right),
\end{equation}
valid to all orders. 

\section{Conclusions}

In this paper, we studied the building blocks of tree-level open and closed string amplitudes on $AdS$. These are given by two infinite towers of world-sheet integrals, $J_w(s,t)$ and $I_w(s,t)$, labelled by words $w$ in the $\{ 0,1 \}$ alphabet, generalising the Euler and complex beta functions respectively. Our main result is a precise relation between the two sets of integrals, of the form
\begin{equation} 
I_w(s,t) = \sum_{w_1,w_2} J_{w_1}(s,t) K^{w_1,w_2}_w(s,t) J_{w_2}(s,t),
\end{equation}
which provides stringy KLT relations on $AdS$. So far, double-copy relations in $AdS$ had been restricted to the supergravity regime, see for instance \cite{Armstrong:2020woi,Albayrak:2020fyp,Zhou:2021gnu,Cheung:2022pdk,Herderschee:2022ntr}. We have found that the inverse of the $AdS$ Kernel has a remarkably simple form, see (\ref{Kinverse}). This structure for the inverse Kernel is reminiscent of what was found in flat space \cite{Mizera:2016jhj,Mizera:2017cqs}. There it was shown that intersection theory provides a geometric interpretation of the KLT relations. In particular, the inverse of the KLT Kernel in flat space can be interpreted as intersection numbers of twisted cycles. It would be very interesting to extend this construction to $AdS$ and find the corresponding underlying twisted intersection theory.    

The integrals considered in this paper generalise the Euler and complex beta functions, and satisfy generalisations of their properties, functional equations and relations. In \cite{Drummond:2013vz,Broedel:2013aza,Stieberger:2013wea} it has been shown that the Drinfeld and Deligne associators are the generating functions of the Euler and complex beta functions respectively, or more generally open and closed string amplitudes in flat space. It would be very interesting to generalise this construction for the higher-weight integrals considered in this paper. 

Another interesting direction is the generalisation to higher-multiplicity amplitudes. The first step is to identify the corresponding building blocks.  Presumably, these involve the insertions of multiple polylogarithms in two or more variables for open strings and their single-valued analogs for closed strings. See \cite{DelDuca:2016lad,Frost:2023stm,Frost:2025lre}, where such single-valued functions are introduced and studied. 

In this work, we have focused on the relation between building blocks for open and closed string amplitudes on $AdS$. Now one can focus on specific amplitudes in specific theories, and look for KLT/double-copy relations, extending the beautiful work in flat space, see \cite{Kawai:1985xq,Bern:2010ue,Bern:2019prr}, to $AdS$.  Our results allow to compute in an analytic/closed form the first $AdS$ curvature corrections for all results available in the literature. An advantage of our expressions - as opposed to their integral form - is that they can be analytically continued away from the region of convergence of the integrals. This opens up the possibility of studying the amplitudes in the complex $s,t$ plane more directly. 

We have developed a new machinery to compute certain classes of integrals over the Riemann sphere. Exact examples in the literature include single-valued hypergeometric functions  and generalisations of those, see for example \cite{Brown:2019jng,Vanhove:2020qtt,Duhr:2023bku,Maldacena:2001km}. Our results complement those, and can be useful in other contexts involving world-sheet computations. 

In this paper, we have shown that much of the beautiful mathematical structure found in string theory amplitudes in flat space persists when $AdS$ curvature corrections are taken into account. It would be interesting to understand the reasons and implications of this for string theory on curved backgrounds.

\acknowledgments
We would like to thank O. Schlotterer and S. Stieberger for useful discussions and comments on the draft. The work of L.F.A. is partially supported by the STFC grant ST/T000864/1. For the purpose of open access, the authors have applied a CC BY public copyright licence to any Author Accepted Manuscript (AAM) version arising from this submission.

\appendix

\section{Conventions}

\label{conventions}

We define multiple zeta values as
\begin{equation}
\zeta(s_1,\cdots,s_d) = \sum^\infty_{\ell_1>\ell_2>\cdots>\ell_d>0} \frac{1}{\ell_1^{s_1}\ell_2^{s_2} \cdots \ell_d^{s_d}},
\end{equation}
where $d$ is called the depth and $s_1+\cdots+s_d$ is called the weight. Multiple zeta values are intimately connected to MPLs \cite{Remiddi:1999ew}. These are functions $L_w(x)$ of one variable, labelled by a word $w$ in the alphabet with letters $\{0, 1\}$. They are recursively defined by the relations
\begin{equation}\label{eq:DerivativeOfMPLs}
\frac{d}{dx} L_{a w}(x) = \frac{1}{x-a} L_{w}(x),~~~a=0,1,
\end{equation}
together with $ L_{e}(x)=1$ for the empty word and $\lim_{x \to 0} L_w(x)=0$, unless $w=0^p$ in which case $L_{0^p}(x) = \frac{1}{p!}\log^p x$. Multiple zeta values can also be defined as multiple polylogarithms evaluated at $x=1$. In our conventions
\begin{equation}
\zeta(s_1,\cdots,s_d)=(-1)^d L_{0^{s_1-1}1 0^{s_2-1} \cdots 0^{s_d-1}1}(1),
\end{equation}
with the regularisation condition $L_{1^p}(1)=0$. Note that also $L_{0^p}(1)=0$. Polylogarithms satisfy shuffle relations
\begin{equation}\label{eq:ShuffleRelationsOfMPLs}
L_w(x) L_{w'}(x) = \sum_{W \in w \shuffle w'} L_W(x).
\end{equation}
Evaluating these at $x=1$, we obtain shuffle relations for the multiple zeta values. Polylogarithms can also be written as iterated integrals, which make manifest their differential relations
\begin{equation}
L_{a_1a_2\cdots a_r}(x) =\int_0^x \frac{dx_1}{x_1-a_1} \int_0^{x_1} \frac{dx_2}{x_2-a_2} \cdots .
\end{equation}
The integrals are convergent provided the last letter is $a_r=1$. Note that the shuffle relations can be used to write all MPLs in terms of these and powers of $L_0(x)=\log x$, so we will often restrict to this case, without loss of generality. It is often convenient to introduce a condensed notation:
\begin{equation}
L_{s_1,\cdots,s_d}(x) = (-1)^d L_{0^{s_1-1}1 0^{s_2-1} \cdots 0^{s_d-1}1}(x),
\end{equation}
with $s_i >0$, so that $\zeta(s_1,\cdots,s_d)=L_{s_1,\cdots,s_d}(1)$. In compact notation,
\begin{eqnarray}
\frac{d}{dx} L_{s_1,\cdots,s_d}(x)  &=& \frac{1}{x} L_{s_1-1,\cdots,s_d}(x)~~~\text{for $s_1>1$},\\
\frac{d}{dx} L_{1,\cdots,s_d}(x) &=& \frac{1}{1-x}  L_{s_2,\cdots,s_d}(x).
\end{eqnarray}
For all $s_i >0$, we can write down a regular series expansion around zero
\begin{equation}
L_{s_1,\cdots,s_d}(x) = \sum^\infty_{\ell_1>\ell_2>\cdots>\ell_d>0} \frac{x^{\ell_1}}{\ell_1^{s_1}\ell_2^{s_2} \cdots \ell_d^{s_d}}.
\end{equation}
It is convenient to introduce non-commutative variables $\{ e_0,e_1 \}$, associated to the letters $\{ 0,1 \}$, and the generating function
\begin{equation}
L(e_0,e_1;x) = L_e(x)+ L_0(x)e_0 + L_1(x)e_1 + L_{00}(x)e_0^2 + L_{01}(x)e_0 e_1+ \cdots,
\end{equation}
so that as a consequence of the derivative relations
\begin{equation}
\frac{\partial }{\partial x} L(e_0,e_1;x) = \left(\frac{e_0}{x}+\frac{e_1}{x-1} \right) L(e_0,e_1;x),
\end{equation}
with the boundary condition $L(e_0,e_1;x)  = e^{e_0 \log x}$ as $x \to 0$. The generating function for multiple zeta values, or in other words the regularised generating function $L(e_0,e_1;x)$ at $x=1$, is the Drinfeld associator
\begin{equation}
Z(e_0,e_1)=L(e_0,e_1;1),
\end{equation}
which satisfies the duality relation
\begin{equation}
Z(e_0,e_1)Z(e_1,e_0)=1.
\end{equation}
As a series expansion in $e_0,e_1$, it is given by
\begin{equation}
Z(e_0,e_1)=1-\zeta(2) [e_0,e_1]+\zeta(3)[[e_0,e_1],e_0+e_1]+ \cdots,
\end{equation} 
where $ [e_0,e_1]=e_0 e_1 - e_1 e_0$ denotes the commutator. The Drinfeld associator can also be defined by $Z(e_0,e_1)=L(e_1,e_0;1-x)^{-1}L(e_0,e_1;x)$. MPLs are multi-valued functions when extended to the complex plane. It is possible to define single-valued functions ${\cal L}_w(z)$, denoted single-valued multiple polylogarithmns (SVMPLs). These are labelled by a word $w$ in the alphabet with letters $\{0, 1\}$ such that they satisfy the same derivative relations as MPLs, under holomorphic derivatives  \cite{Brown:2004ugm}
\begin{equation}
\frac{\partial }{\partial z} {\cal L}_{a w}(z) = \frac{1}{z-a} {\cal L}_{w}(z),~~~a=0,1,
\end{equation}
together with $ {\cal L}_{e}(z)=1$ for the empty word and $\lim_{z \to 0} {\cal L}_w(z)=0$, unless $w=0^p$ in which case ${\cal L}_{0^p}(z) = \frac{1}{p!}\log^p|z|^2$. SVMPLs satisfy the shuffle identities
\begin{equation}
{\cal L}_w(z) {\cal L}_{w'}(z) = \sum_{W \in w \shuffle w'} {\cal L}_W(z).
\end{equation}
Evaluated at one, SVMPLs generate single-valued multiple zeta values
\begin{equation}
\zeta^{sv}(s_1,\cdots,s_d)={\cal L}_{s_1,\cdots,s_d}(1).
\end{equation}
More generally, we can define a single valued map such that
\begin{equation}
{\cal L}_w(z) = sv\left(L_w(z)  \right).
\end{equation}
As for MPLs, we can define a generating function
\begin{equation}
{\cal L}(e_0,e_1;z) = {\cal L}_e(z)+ {\cal L}_0(z)e_0 + {\cal L}_1(z)e_1 +{\cal L}_{00}(z)e_0^2 + {\cal L}_{01}(z)e_0 e_1+ \cdots
\end{equation}
so that as a consequence of the derivative relations
\begin{equation}
\frac{\partial }{\partial z} {\cal L}(e_0,e_1;z) = \left(\frac{e_0}{z}+\frac{e_1}{z-1} \right) {\cal L}(e_0,e_1;z), 
\end{equation}
with the boundary condition ${\cal L}(e_0,e_1;z)  = e^{e_0 \log |z|^2}$ as $z \to 0$. Its regularised value at $z=1$ defines the Deligne associator 
\begin{equation}
W(e_0,e_1)={\cal L}(e_0,e_1;1),
\end{equation}
which satisfies the duality relation
\begin{equation}
W(e_0,e_1)W(e_1,e_0)=1.
\end{equation}
In an expansion, the Deligne associator is given by
\begin{equation}
W(e_0,e_1)=1 + 2 \zeta(3)[[e_0,e_1],e_0+e_1]+\cdots .
\end{equation}
The Deligne associator can also be defined by $W(e_0,e_1)={\cal L}(e_1,e_0;1-z)^{-1}{\cal L}(e_0,e_1;z)$. SVMPLs can be constructed from MPLs in a recursive way \cite{FrancisB}. We have
\begin{equation}\label{eq:ConstructionOfSVMPLsFromMPLs}
{\cal L}(e_0,e_1;z) = L(e_0,e_1;z)  L^R(e_0,e'_1;\bar z),
\end{equation}
where the Reverse operator $R$ acts by reversing the order of the words labelling the components of ${\cal L}(e_0,e'_1;\bar z)$
\begin{equation}
 L^R(e_0,e'_1;z)=1 + \cdots + L_{0011}(z) e'_1 e'_1 e_0 e_0 + \cdots
\end{equation}
and $e_1'$ is a deformed variable which can be obtained from $e_0,e_1$ by solving
\begin{equation}
Z^R(e_0,e'_1) e'_1 (Z^R(e_0,e'_1))^{-1} =Z(e_0,e_1)^{-1} e_1 Z(e_0,e_1)
\end{equation}
recursively. In an expansion, we obtain
\begin{equation}
e'_1 = e_1 -2 \zeta(3)[e_0+e_1,[e_1,[e_0,e_1]]]+ \cdots.
\end{equation}
The above construction implies that the Drinfeld and Deligne associators are related by
\begin{equation}
W(e_0,e_1)=Z(e_0,e_1) Z(e_0,e_1')^R.
\end{equation}
Furthermore, the generating function ${\cal L}(e_0,e_1;z)$ satisfies the anti-holomorphic KZ equation
\begin{equation}
\frac{\partial}{\partial \bar z} {\cal L}(e_0,e_1;z)  ={\cal L}(e_0,e_1;z)  \left(\frac{e_0}{\bar z}+\frac{e'_1}{\bar z-1} \right).
\end{equation}

\section{Pole structure}

\label{poles}
For applications to dispersive sum rules, it is convenient to have an expansion in poles for the building blocks $J_w(s,t)$. We can always use the shuffle relations to write polylogarithms in terms of $L_0(x)$ and polylogarithms whose label ends in the letter 1. In terms of the $J_w(s,t)$ integrals, this means we can always focus on words ending in $1$ and derivatives of those functions w.r.t. $s$. Let us then use the compact notation. We have
\begin{equation}
J_{s_1,\cdots,s_d}(s,t) = \sum^\infty_{\ell_1>\ell_2>\cdots>\ell_d>0} \frac{\beta(s+\ell_1,t)}{\ell_1^{s_1}\ell_2^{s_2} \cdots \ell_d^{s_d}}.
\end{equation}
This has poles at $s=-n$ and $t=-n$. We can actually write
\begin{equation}
\beta(s,t)=\sum_{n=0}^\infty \frac{(-1)^n \Gamma(t)}{\Gamma(n+1)\Gamma(t-n)} \frac{1}{s+n}=\sum_{n=0}^\infty \frac{(-1)^n \Gamma(s)}{\Gamma(n+1)\Gamma(s-n)} \frac{1}{t+n}.
\end{equation}
For the poles in $t$, we find
\begin{equation}
J_{s_1,\cdots,s_d}(s,t) =\sum_{n=0}^\infty \frac{1}{t+n} \sum^\infty_{\ell_1>\ell_2>\cdots>\ell_d>0} \frac{(-1)^n \Gamma(s+\ell_1)}{\Gamma(n+1)\Gamma(s+\ell_1-n) \ell_1^{s_1}\ell_2^{s_2} \cdots \ell_d^{s_d}}.
\end{equation}
For the poles in $s$, we find
\begin{equation}
J_{s_1,\cdots,s_d}(s,t) =\sum_{n=0}^\infty\sum^\infty_{\ell_1>\ell_2>\cdots>\ell_d>0} \frac{(-1)^n \Gamma(t)}{\Gamma(n+1)\Gamma(t-n)}\frac{1}{\ell_1^{s_1}\ell_2^{s_2} \cdots \ell_d^{s_d}} \frac{1}{s+n+\ell_1}.
\end{equation}
Renaming $n+\ell_1=n'$ so that $\ell_1 \leq n'$, we find
\begin{equation}
J_{s_1,\cdots,s_d}(s,t) =\sum_{n'=0}^\infty  \frac{1}{s+n'}  \sum^{n'}_{\ell_1>\ell_2>\cdots>\ell_d>0} \frac{(-1)^{n'-\ell_1} \Gamma(t)}{\Gamma(n' - \ell_1 +1)\Gamma(t+\ell_1-n')}\frac{1}{\ell_1^{s_1}\ell_2^{s_2} \cdots \ell_d^{s_d}}.
\end{equation}
\section{Poincaré duality}
\label{poincare}
In this appendix, we generalise the relation (\ref{Poincare}) due to Poincaré duality. We consider
\begin{equation}
P(s,t;e_0,e_1) \equiv {\cal J}(s,t;e_0,e_1) {\cal K}(s,t;e_0,e_1) {\cal J}^{R}(-s,-t;-e_0,-e_1) 
\end{equation}
and deduce properties of the function $P(s,t;e_0,e_1)$. In particular, let us consider shifts in $s$
\begin{equation}
P(s+1,t;e_0,e_1) = {\cal J}(s+1,t;e_0,e_1) {\cal K}(s+1,t;e_0,e_1) {\cal J}^{R}(-s-1,-t;-e_0,-e_1).
\end{equation}
Let's recall the shift relations (\ref{shiftJ})
\begin{equation}
{\cal J}(s+1,t; e_0,e_1) =\frac{1}{s+t+e_0+e_1} (s+e_0) {\cal J}(s,t; e_0,e_1)
\end{equation}
and recall that $ {\cal K}(s,t;e_0,e_1)$ is periodic. This leads to 
\begin{equation}
P(s+1,t;e_0,e_1) =\frac{1}{s+t+e_0+e_1} (s+e_0)  P(s,t;e_0,e_1) (s+t+1+e_0+e_1) \frac{1}{s+1+e_0}.
\end{equation}
We can write both shifts in $s$ and $t$ as
\begin{eqnarray*}
(s+t+e_0+e_1)P(s+1,t;e_0,e_1)(s+1+e_0) &=&(s+e_0)  P(s,t;e_0,e_1) (s+t+1+e_0+e_1),\\
(s+t+e_0+e_1)P(s,t+1;e_0,e_1)(t+1+e_1) &=&(t+e_1)  P(s,t;e_0,e_1) (s+t+1+e_0+e_1).
\end{eqnarray*}
We also have symmetry under the exchange of $s$ and $t$, see (\ref{Jsym}) and (\ref{Ksym}). It follows\footnote{We have used the fact that $L(e_0,e_1,x)^{-1}=L^R(-e_0,-e_1,x)$, see \cite{Brown:2013gia}. In particular evaluating this at $x=1$ implies a relation between the Drinfeld associator with {\it reversed} non-commutative variables and $e_i \to -e_i$. }
\begin{eqnarray*}
P(s,t;e_0,e_1)=P(t,s;e_1,e_0).
\end{eqnarray*}
Using this symmetry together with the recursion relations and the result at zeroth order, we obtain
\begin{equation}
{\cal J}(s,t;e_0,e_1) {\cal K}(s,t;e_0,e_1) {\cal J}^{R}(-s,-t;-e_0,-e_1) = -\frac{1}{s+e_0} - \frac{1}{t+e_1}.
\end{equation}

\bibliography{refholo} 
\bibliographystyle{utphys}

\end{document}